**Enumeration of Hamiltonian walks in two and three dimensional lattices**

K. Silpaja Chandrasekar[1] and M.V.Sangaranarayanan*

Department of Chemistry, Indian Institute of Technology - Madras

Chennai -600036 India

[1] E-mail: silpajachandrasekar@gmail.com

* E-mail: sangara@iitm.ac.in

Abstract

We report an efficient methodology for enumerating the Hamiltonian walks in two and three dimensional lattices of large sizes, using the concept of centroids. This strategy, with the help of JAVA programming enables the exact computation of the Hamiltonian walks for square and simple cubic lattices. These estimates are useful in designing the protein sequences using hydrophobic-polar lattice models as well as in the analysis of secondary structures in compact polymers.

---

## 1. Introduction

Hamiltonian walks in two and three dimensional lattices refer to walks wherein every site is visited only once without any intersecting paths and the exact enumeration of these walks is rendered non-trivial in view of the computational complexity [1]. These values provide insights into compact conformations of proteins in so far as the ground state can be deciphered after assigning appropriate pairwise interaction energies between the hydrophobic and polar residues within the lattice model framework [2]. Despite their simplistic nature, the lattice models have provided significant insights into the designing of proteins. Furthermore, the Hamiltonian walks in lattices may serve as simple models for protein conformations [3].

In the case of compact polymers, the Hamiltonian walks refer to the self-avoiding walk wherein all the lattice sites are visited [4]and Monte Carlo simulations have been performed for square lattices of various sizes using 'transition graphs' [5] for studying the scaling behaviour of polymers. As mentioned elsewhere [6, 7], it is 'computationally infeasible' to estimate the Hamiltonian walks for large lattices.

There exist different variants of the Hamiltonian walks such as Hamiltonian cycles [8]. If one of the edges is removed from the cycles, the latter leads to the Hamiltonian walks. Alternately, if the endpoints are adjacent, the Hamiltonian walks yield cycles. As pointed out elsewhere [8], Hamiltonian cycle problem is NP complete and is a particular case of the travelling salesman problem. A comprehensive approach to the study of Hamiltonian paths and cycles is that due to Jacobsen [1] wherein square and cubic lattices of varying sizes were analysed. It is of interest to point out a recent study wherein closed Hamiltonian walks on fractal lattices enable the estimation of the partition function and persistence length [9]. Here we provide a novel method of computing the Hamiltonian walks for square and simple cubic lattices with the help of JAVA programming, using the concept of centroids.

## 2. Methodology

### 2.1 Creation of the lattices

Initially, the algorithm involves the grouping various sites into different clusters. In the clustering problem, the matrix elements associated with the sites are designated as x (1)... x (n), and are grouped into a few clusters using the calculation of the centroid [10].The first part of this algorithm is to predict the centroids $c(i)$ for each data point in the matrix. The clustering algorithm involves the following steps: (i) each element of the matrix corresponds to a particular starting point of a chosen lattice arrangement. To avoid revisiting, the classification of the same structures with identical starting points is carried out using the connectivity between two sites. (ii) The partition of the elements is accomplished by identifying the clusters having the same starting point. (iii)The formation of the clusters continues until no data point is left. The second part of the algorithm keeps track of the sites that form a cluster. The initial and destination points are updated every time after a complete lattice is formed. When all the inner lattice points are occupied and the last site gets connected, each point is checked in order to verify the exclusion of identical lattices.

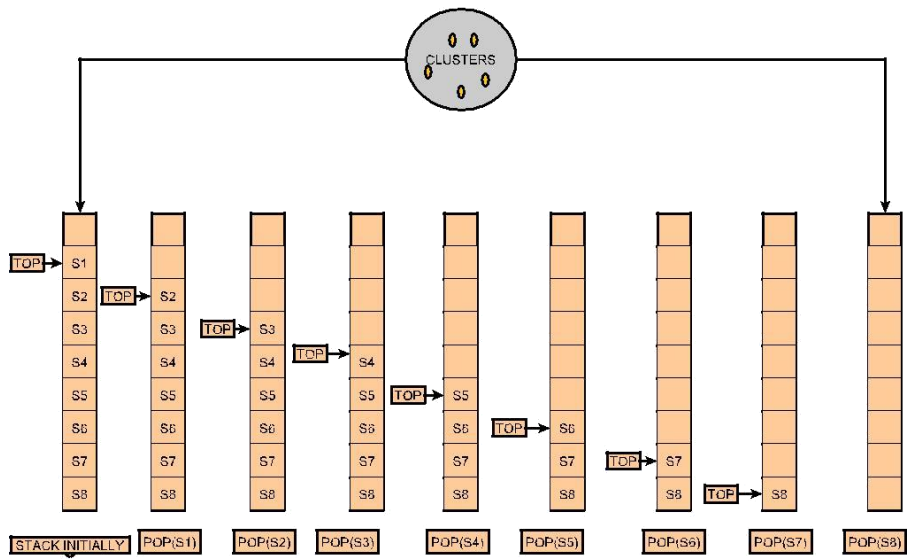

Figure 1: Representation of the methodology for computing the Hamiltonian walks using stacks

The operation performed at each stack is designated as V.

$$V = \sum_{i=1}^{n} \sum_{x=Si}^{n} (x - \mu_i)^2$$

(1)

The centroid is computed for all values of i ranging from 1 to n and x is the Euclidean distance obtained using

$$x = \|x_i - x_j\|^2$$

(2)

The computation of the centroids $C^i$ is performed for all i =1…. 2n. The points are being clustered using their distances from the centroid and i denotes the intensity of the ith centroid. [10]

$$C^i = \min(x - \mu_i)^2$$

$$\mu_i = \frac{\sum_{i+1}^{n} 1 \{C^i = j\} x^i}{\sum_{i+1}^{n} 1 \{C^i = j\}}$$

### 2.2.1 Operations in stack

After the formation of the cluster, the stack operations such as insertion and deletion of elements are performed to track the visit [11]. The centroid for the particular cluster is the starting point. The top stack value corresponds to the centroid of the particular cluster. Each site inserted into a stack is followed by the starting point. All the sites perform push and pop operations till an empty stack results. The stack operations are carried out such that the recurrence of any data structure is avoided. A stack is either empty or it contains elements (S1 to Sn). The matrices with the corresponding number of sites are stored in the form of stacks with indices wherein an index corresponds to a memory address. The sites are arranged such that the inner surface points of the stack has a lower index than the site present at the top. The points at the inner surface represent all the elements except the starting point. Since all the structures have reflection and symmetry properties, identical configurations can be removed. For example, if the surface exists with multiple structures, then the reflected structures will also have the same configuration with the identical number of structures. The algorithm enables the push and pop operations to be performed for all the elements in the stack, in the case of two and three dimensional lattices. At each stage, the number of lattices formed by performing the push and pop operations has been systematically tracked. The processing of these possibilities for the formation of lattices is continued until each site becomes a starting point.

### 2.2.2 Algorithm

**Square lattice**

The algorithm for the enumeration of a square lattice spanning 1 to 9 data points is:(i)The data point is placed at the top of the stack. This configuration has the signature($3^2$), which enters the stack with a count of 1.(ii)A centroid is calculated with all the data points by getting the position of all points of a single cluster followed by summation and then division by the number of points.(iii)The next step involves finding the centroid of the matrix c (i) with all the data points in the matrix where i ranges from 1 to 9. The site must occupy the position of stacks at position W-j where W is 9 and j is the data point denoting the occupied site.(iv)The first data point is placed in the last column until a new site is occupied.(v)The last data point is placed in the first column, followed by visiting all the points. If a new site is occupied, we ensure that the data point is marked to ensure that it has been visited once. Consequently, the sites will get updated and there will no revisiting of the data point.(vi)This is repeated through all existing data points until n =9 while no data point should be inserted again in the stack and (vii)If the new stack is occupied, we ensure that the data point is marked until the lower border.

**Simple cubic lattice**

The steps involved in the enumeration of a simple cubic lattice spanning typically 1 to 8 data points is as follows: (i)Initially,the data point is placed at the top of the stack. This configuration has the signature$(2^3)$, which enters the stack with a count of 1.(ii)A centroid is calculated with all the data points by getting the position of all points of a single cluster, summing all of them and dividing by the number of points.(iii)We then find the centroid of the matrix c (i) with all the data points in the matrix where i ranges from 1 to 8. The site must occupy the position of stacks at position W-j where W is 8 and j is the data point getting occupied.(iv)The first data point is placed in the last column until a new site is occupied; then we mark the signature as having touched the lower stack.(v)subsequently, the last data point is placed in the first column and systematically all the data points are visited. If a new site is occupied, we ensure that the data point is marked and it has been visited once. With this, the site model will be updated and there will no revisiting of the data point.(vi)This procedure is repeated through all data points until n =8. Again, no data point should be inserted in the stack and (vii)If the new stack is occupied, we ensure that the data point is marked until the lower border.

This algorithm commences at the level of i-1 and continues recursively through all the way to arrays pointing deeper into the values. Firstly, the pointer points to the memory address having the value corresponding to the starting point of the matrix. Then, it is pushed out and the second element is brought to the top. The pop operation takes place until all the symmetry operations are carried out. When this operation is performed until the end of the stack and with the condition that each site is visited only once, a complete square lattice or a simple cubic lattice is formed. A loop performs the connection from surface point S1 to Sn. If one combines two elements 1 and n, with the connections S1.Sn where S1 and Sn correspond to the starting and ending point in a lattice, the new stack structure is created.

At the initial step, the entire stack is empty, devoid of any assigned values. After the loop is executed at each step, the configuration is updated and configured, the stack becoming empty at the final step. In order for the condition that each site should be visited only once, the lattice forms a complete structure due to the connectivity of all the sites.

The first level of the stack is an array of S1 to S8. The value of each element in this array, possesses the memory address of another similar array structure. When the index value is changed to the next place in the stack, it implies that the site has been visited once. This process continues recursively until the array pointer reaches the end of the stack and when all sites have been visited once. Thus, at each level, the array values will have no pointers to point out the array from the starting to end array in the memory, but instead the value alone.

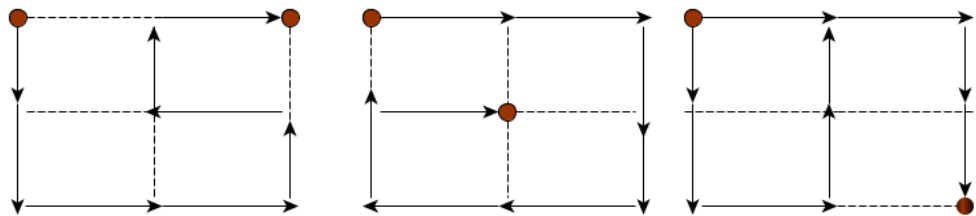

Figure 2: Typical Hamiltonian walks for a square lattice of 9 sites, denoting the destination sites

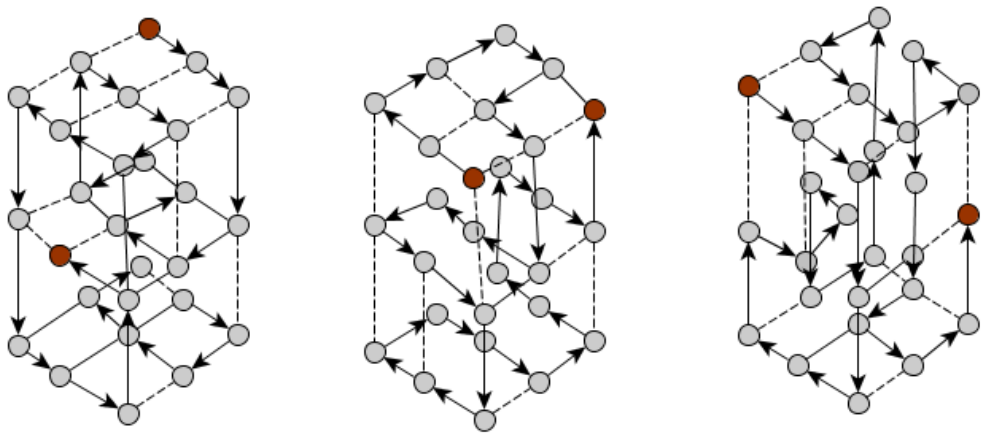

Figure 3: Typical Hamiltonian walks for a cubic lattice of 27 sites, denoting the destination sites

The final step shows the structure of a complete square and cubic lattice by the two ends connected with all the intermediate sites also connected together.

3. Results and Discussion

**(A) Square lattices**

Tables 1 and 2 provide the Hamiltonian walks for square and cubic lattices respectively. We have enumerated these by the clustering algorithm till $N = 19^2$ for square and $N = 6^3$ for cubic lattices and these values were generated within *a few seconds* using JAVA programming. In the case of square lattices, the exact values exist till $N = 17^2$ [12] by transfer matrix algorithm and the values of Table 1 are entirely in agreement with Table 6 of [12]. Furthermore, the compact self-avoiding walks have also been enumerated for *rectangular strips* of square lattices, using the method of generating functions [13].

**(B) Cubic lattices**

For the cubic lattices designated as 3x3x4 and 3x4x4 here, the values obtained here are identical with those reported by Pande et al [7], after multiplication by 8, in view of excluding the symmetrical configurations therein. The results of Table 2 are consistent with the hitherto-known results till $4^3$ (Table 1 of [3] and for lattices larger than these, no exact enumerations are available. The value of 2480304 (for 3x3x3) and 677849536 for (3x3x4) is also identical with the estimated number of *Hamiltonian chains* of order 1 in Table 4 of [1]. Furthermore, a linear relation between logarithm of the number of Hamiltonian walks with the lattice sites has been demonstrated [7], consistent with the asymptotic limit of the Flory theory. On account of the restriction on memory and byte distribution, Hamiltonian walks for lattices larger than $6^3$ could not be estimated by us.

Although asymptotic analysis using growth constants to predict the Hamiltonian walks exist [14], it is gratifying to note that exact computations are feasible for moderately large lattices. The present analysis can be extended *mutatis mutandis* to develop new algorithms for the designing of protein sequences [15, 16] and elucidation of the compact structures of secondary polymers. Elsewhere [17], we have demonstrated a simple methodology of deducing protein conformations of two and three dimensional lattice models using JAVA programming. The JAVA programming in conjunction with the concept of centroids has enabled the analysis for large lattices – hitherto-considered almost impossible.

## 4. Conclusions

We have enumerated the Hamiltonian walks in square (LxL) and cubic (L1xL2xL3) lattices using JAVA programming, in conjunction with the concept of centroids. The estimated values are in agreement with the hitherto-known results for smaller lattice sizes while exact computations for larger lattices are provided for the first time. The significance of the methodology in the design of protein sequences is indicated.

## Acknowledgements

This work was supported by the SERB, Department of Science and Technology (Grant number MTR/2017/000705).

Table 1: Estimated Hamiltonian walks for square lattices for LxL sites

| L | Number of Hamiltonian walks | L | Number of Hamiltonian walks |
|---|---|---|---|
| 4 | 104 | 12 | 1.73376E+22 |
| 5 | 10180 | 13 | 1.99892E+27 |
| 6 | 111712 | 14 | 4.62865E+30 |
| 7 | 67590888 | 15 | 2.93744E+36 |
| 8 | 2688307514 | 16 | 2.74788E+40 |
| 9 | 9.62877E+12 | 17 | 9.45551E+46 |
| 10 | 1.44578E+15 | 18 | 3.45011E+59 |
| 11 | 2.97259E+19 | 19 | 4.48575562421E+76 |

Table 2: Estimated Hamiltonian walks for cubic lattices of L1xL2xL3 sites

| L1xL2xL3 | Number of Hamiltonian walks | L1xL2xL3 | Number of Hamiltonian walks |
|---|---|---|---|
| 2 x 2 x 2 | 72 | 4 x 3 x 5 | 1.96E+13 |
| 2 x 2 x 3 | 584 | 4 x 3 x 3 | 6.78E+21 |
| 2 x 3 x 3 | 16880 | 5 x 5 x 5 | 9.98E+29 |
| 3 x 3 x 3 | 2480304 | 6 x 6 x 4 | 6.76E+35 |
| 3 x 3 x 4 | 677849536 | 5 x 5 x 6 | 5.26E+40 |
| 3 x 4 x 4 | 1.07305E+12 | 5 x 6 x 6 | 7.72E+47 |
| 4 x 4 x 4 | 2.77E+16 | 6 x 6 x 6 | 6.59E+70 |

**Appendix A**

Listing of JAVA programming for Hamiltonian walks in square lattices
import java.io.*;
import java.util.*;
import java.lang.*;
import java.text.*;

public class Centroid
{

// finds the mean of the dataset for grouping
public void Grouping(double[] Matrix1, double[] Matrixele2, int centerofele)

```
{
int Noofelem = centerofele;
double[] ClustMatrix1 = new double[centerofele];
double[] ClustMatrixele2 = new double[centerofele];
this.getMeansetCentroid(Matrix1, Matrixele2, centerofele);
DecimalFormat dec = new DecimalFormat("0.00");
for(int i = 0;i<Matrix1.length;i++)
{
String result1 = dec.format(Matrix1[i]);
String result2 = dec.format(Matrixele2[i]);
System.out.println("\n Cords are ( " + result1 + " , " + result2 + ")");
}
//setting random datasets as centroids
for(int i = 0; i<centerofele;i++)
{

ClustMatrix1[i] = Matrix1[i];

ClustMatrixele2[i] = Matrixele2[i];

}

this.groupCordtoCluster(Matrix1,Matrixele2,ClustMatrix1,ClustMatrixele2);
}

public void groupCordtoCluster(double[] Matrix1, double[] Matrixele2, double[]
ClustMatrix1,double[] ClustMatrixele2)
{
double temp ;
int size = Matrix1.length;int clustsize = ClustMatrix1.length;
int clusterComparison = clustsize;
int[] grouping = new int[size - clustsize];
double[] ClustgroupX = new double[size - clustsize];
double[] ClustgroupY = new double[size - clustsize];
int tempint = -1;

//grouping the dataset to respective clusters by comparing the distance
for(int i = clusterComparison; i < size;i++)
{
temp = 0;
for(int j = 0;j<clustsize;j++)
{
if (j == 0)
tempint++;
if(temp == 0)
{
```

```
temp = Math.sqrt(Math.pow((Matrix1[i]-ClustMatrix1[j]),2) +
Math.pow((Matrixele2[i]-ClustMatrixele2[j]),2));
grouping[tempint] = j;
ClustgroupX[tempint] = Matrix1[i];
ClustgroupY[tempint] = Matrixele2[i];
}
else if (temp > Math.sqrt(Math.pow((Matrix1[i]-ClustMatrix1[j]),2) +
Math.pow((Matrixele2[i]-ClustMatrixele2[j]),2)))
{
temp = Math.sqrt(Math.pow((Matrix1[i]-ClustMatrix1[j]),2) +
Math.pow((Matrixele2[i]-ClustMatrixele2[j]),2));
grouping[tempint] = j;
ClustgroupX[tempint] = Matrix1[i];
ClustgroupY[tempint] = Matrixele2[i];
}
}
void showpop(Stack st) {
System.out.print("pop -> ");
Integer a = (Integer) st.pop();
System.out.println(a);
System.out.println("stack: " + st);
}
public void main(String args[]) {
Stack st = new Stack();
System.out.println("stack: " + st);
showpush(st, 42);
showpush(st, 66);
showpush(st, 99);
showpop(st);
showpop(st);
showpop(st);
try {
showpop(st);
} catch (EmptyStackException e) {
System.out.println("empty stack");
}
}
int twoD[][] = new int[4][];
twoD[0] = new int[1];
twoD[1] = new int[2];
twoD[2] = new int[3];
twoD[3] = new int[4];
int i, j, k = 0;
for(i=0; i<4; i++)
for(j=0; j<i+1; j++) {
twoD[i][j] = k;
```

```
k++;
}
for(i=0; i<4; i++) {
for(j=0; j<i+1; j++)
System.out.print(twoD[i][j] + " ");
System.out.println();
}
DecimalFormat dec = new DecimalFormat("0.00");
String result1, result2, result3, result4;
for(int i = 0; i<grouping.length;i++,j++,k++)
{
//for(int i = 1; i< centerofele;i++) {
System.out.println("-----------------------");
System.out.println("Clusters for group " + grouping[i]);
result1 = dec.format(Matrix1[grouping[i]]);
result2 = dec.format(Matrixele2[grouping[i]]);
result3 = dec.format(ClustgroupX[i]);
result4 = dec.format(ClustgroupY[i]);
System.out.println("Cordinates are (" + result1 + " , " + result2 + ")");
System.out.println("-----------------------");
System.out.println("Clusters for group " + grouping[i]);
System.out.println("Cordinates are (" + result3 + " , " + result4 + ")");

}
}
public void getMeansetCentroid(double[] Matrix1, double[] Matrixele2, int
centerofele)
{
double xCord, yCord;
double MAX=0,distance, tempd1, tempd2;
double[] Distances = new double[Matrix1.length];
int reference, i, j, temp1, temp2, point, length;
reference = i = j = temp1 = temp2 = point =0;
int[] centroids;

class Stack<Item> implements Iterable<Item> {
private final PrintStream StdOut = null;
private int N;              // size of the stack
private Node<Item> first;
private Object StdIn;     // top of stack

// helper linked list class
class Node<Item> {
private Item item;
private Node<Item> next;
}
```

```java
/**
 * Initializes an empty stack.
 */
public Stack() {
first = null;
N = 0;
}

/**
 * Is this stack empty?
 * @return true if this stack is empty; false otherwise
 */
public boolean isEmpty() {
return first == null;
}

/**
 * Returns the number of items in the stack.
 * @return the number of items in the stack
 */
public int size() {
return N;
}

/**
 * Adds the item to this stack.
 * @param item the item to add
 */
public void push(Item item) {
Node<Item> oldfirst = first;
first = new Node<Item>();
first.item = item;
first.next = oldfirst;
N++;
}

/**
 * Removes and returns the item most recently added to this stack.
 * @return the item most recently added
 * @throws java.util.NoSuchElementException if this stack is empty
 */
public Item pop() {
if (isEmpty()) throw new NoSuchElementException("Stack underflow");
Item item = first.item;        // save item to return
first = first.next;            // delete first node
```

```
N--;
return item;              // return the saved item
}

/**
 * Returns (but does not remove) the item most recently added to this stack.
 * @return the item most recently added to this stack
 * @throws java.util.NoSuchElementException if this stack is empty
 */
public Item peek() {
if (isEmpty()) throw new NoSuchElementException("Stack underflow");
return first.item;
}

/**
 * Returns a string representation of this stack.
 * @return the sequence of items in the stack in LIFO order, separated by spaces
 */
public String toString() {
StringBuilder s = new StringBuilder();
for (Item item : this)
s.append(item + " ");
return s.toString();
}

public Iterator<Item> iterator() {
return new ListIterator<Item>(first);
}

// an iterator, doesn't implement remove() since it's optional
class ListIterator<Item> implements Iterator<Item> {
private Node<Item> current;

public ListIterator(Node<Item> first) {
current = first;
}
public boolean hasNext()  { return current != null;              }
public void remove()      { throw new UnsupportedOperationException();  }

public Item next() {
if (!hasNext()) throw new NoSuchElementException();
Item item = current.item;
current = current.next;
return item;
}
```

```java
}
public static void main (String args[]) throws IOException
{
Scanner input = new Scanner(System.in);
ClustNumber=input.nextInt();
String [][] numbers = new String [6][2];
double Cordx[] =new double[6];
double Cordy[] =new double[6];
File file = new File("sam.csv");
BufferedReader bufRdr = new BufferedReader(new FileReader(file));
String line = null;
int row = 0;
int col = 0;

//read each line of text file
while((line = bufRdr.readLine()) != null && row< 6 )
{
StringTokenizer st = new StringTokenizer(line,",");
while (st.hasMoreTokens())
{
//get next token and store it in the array
numbers[row][col] = st.nextToken();
col++;
}
col = 0;
row++;
}
for(row=0;row < 6;row++)
{
for(col=0; col<2;col++)
{
System.out.print(" " + numbers[row][col]);
}
System.out.println(" ");}
for(row=0;row<6;row++)
{

Cordx[row]=Double.parseDouble(numbers[row][0]);
Cordy[row]=Double.parseDouble(numbers[row][1]);
//System.out.print(" " + Cordx[row]);
}

for(row=0;row<6;row++)
{
System.out.print(" " + Cordx[row]);
//System.out.print("\n " + Cordy[row]);
```

```java
}
System.out.print(" \n");
for(row=0;row<6;row++)
{

System.out.print(" " + Cordy[row]);
}
cent.Grouping(Cordx,Cordy,ClustNumber);

}
}
class StackDemo {
void showpush(Stack st, int a) {
st.push(new Integer(a));
System.out.println("push(" + a + ")");
System.out.println("stack: " + st);
}

/**
* Unit tests the <tt>Stack</tt> data type.
*/
public void main(String[] args) {
Stack<String> s = new Stack<String>();
while (!((String) StdIn).isEmpty()) {
String item = StdIn.toString();
if (!item.equals("-")) s.push(item);
else if (!s.isEmpty()) StdOut.print(s.pop() + " ");
}
StdOut.println("(" + s.size() + " left on stack)");
}
}
for(j = 1; j < Matrix1.length;j++)
{
Distances[j-1] = Math.sqrt(Math.pow((Matrix1[j]-Matrix1[0]),2) +
Math.pow((Matrixele2[j]-Matrixele2[0]),2));
}

for(i=0;i<Matrix1.length-1;i++)
{
for(j=0;j<Matrix1.length-1-i;j++)
{
if(Distances[j+1] < Distances[j])
{
distance = Distances[j];
tempd1 = Matrix1[j];
```

```
tempd2 = Matrixele2[j];
Distances[j] = Distances[j+1];
Matrix1[j] = Matrix1[j+1];
Matrixele2[j] = Matrixele2[j+1];
Distances[j+1] = distance;
Matrix1[j+1] = tempd1;
Matrixele2[j+1] = tempd2;
}
}
}
//recalculation of centroids
point = Matrix1.length;
do
{
if(Matrix1.length % centerofele != 0)
point--;
}while(point % centerofele != 0);
length = point/centerofele;
for(i=0;i<Matrix1.length;i=length+i)
{
if((i+length-1) > point)
break;
tempd1 = Matrix1[i];
tempd2 = Matrixele2[i];
Matrix1[i] = Matrix1[i+length-1];
Matrixele2[i] = Matrixele2[i+length-1];
Matrix1[i+length-1] = tempd1;
Matrixele2[i+length-1] = tempd2;
}
}
}
```

**Appendix B**

<u>Listing of JAVA programming for Hamiltonian walks in cubic lattices</u>

```
import java.io.*;
import java.util.*;
import java.lang.*;
import java.text.*;

public class Centroid
{
```

```
// finds the mean of the dataset for grouping
public void Grouping(double[] Matrix1, double[] Matrixele2, int centerofele)
{
int Noofelem = centerofele;
double[] ClustMatrix1 = new double[centerofele];
double[] ClustMatrixele2 = new double[centerofele];
this.getMeansetCentroid(Matrix1, Matrixele2, centerofele);
DecimalFormat dec = new DecimalFormat("0.00");
for(int i = 0;i<Matrix1.length;i++)
{
String result1 = dec.format(Matrix1[i]);
String result2 = dec.format(Matrixele2[i]);
System.out.println("\n Cords are ( " + result1 + " , " + result2 + ")");
}
//setting random datasets as centroids
for(int i = 0; i<centerofele;i++)
{

ClustMatrix1[i] = Matrix1[i];

ClustMatrixele2[i] = Matrixele2[i];

}

this.groupCordtoCluster(Matrix1,Matrixele2,ClustMatrix1,ClustMatrixele2);
}

public void groupCordtoCluster(double[] Matrix1, double[] Matrixele2, double[]
ClustMatrix1,double[] ClustMatrixele2)
{
double temp ;
int size = Matrix1.length;int clustsize = ClustMatrix1.length;
int clusterComparison = clustsize;
int[] grouping = new int[size - clustsize];
double[] ClustgroupX = new double[size - clustsize];
double[] ClustgroupY = new double[size - clustsize];
int tempint = -1;

//grouping the dataset to respective clusters by comparing the distance
for(int i = clusterComparison; i < size;i++)
{
temp = 0;
for(int j = 0;j<clustsize;j++)
{
if (j == 0)
```

```
tempint++;
if(temp == 0)
{
temp = Math.sqrt(Math.pow((Matrix1[i]-ClustMatrix1[j]),2) +
Math.pow((Matrixele2[i]-ClustMatrixele2[j]),2));
grouping[tempint] = j;
ClustgroupX[tempint] = Matrix1[i];
ClustgroupY[tempint] = Matrixele2[i];
}
else if (temp > Math.sqrt(Math.pow((Matrix1[i]-ClustMatrix1[j]),2) +
Math.pow((Matrixele2[i]-ClustMatrixele2[j]),2)))
{
temp = Math.sqrt(Math.pow((Matrix1[i]-ClustMatrix1[j]),2) +
Math.pow((Matrixele2[i]-ClustMatrixele2[j]),2));
grouping[tempint] = j;
ClustgroupX[tempint] = Matrix1[i];
ClustgroupY[tempint] = Matrixele2[i];
}
}
void showpop(Stack st) {
System.out.print("pop -> ");
Integer a = (Integer) st.pop();
System.out.println(a);
System.out.println("stack: " + st);
}
public void main(String args[]) {
Stack st = new Stack();
System.out.println("stack: " + st);
showpush(st, 42);
showpush(st, 66);
showpush(st, 99);
showpop(st);
showpop(st);
showpop(st);
try {
showpop(st);
} catch (EmptyStackException e) {
System.out.println("empty stack");
}
}
int threeD[][][] = new int[3][4][5];
int i, j, k;
for(i=0; i<3; i++)
for(j=0; j<4; j++)
for(k=0; k<5; k++)
threeD[i][j][k] = i * j * k;
```

```
for(i=0; i<3; i++) {
for(j=0; j<4; j++) {
for(k=0; k<5; k++)
System.out.print(threeD[i][j][k] + " ");
System.out.println();
}
System.out.println();
}
DecimalFormat dec = new DecimalFormat("0.00");
String result1, result2, result3, result4;
for(int i = 0; i<grouping.length;i++,j++,k++)
{
//for(int i = 1; i< centerofele;i++) {
System.out.println("----------------------");
System.out.println("Clusters for group " + grouping[i]);
result1 = dec.format(Matrix1[grouping[i]]);
result2 = dec.format(Matrixele2[grouping[i]]);
result3 = dec.format(ClustgroupX[i]);
result4 = dec.format(ClustgroupY[i]);
System.out.println("Cordinates are (" + result1 + " , " + result2 + ")");
System.out.println("----------------------");
System.out.println("Clusters for group " + grouping[i]);
System.out.println("Cordinates are (" + result3 + " , " + result4 + ")");

}
}
public void getMeansetCentroid(double[] Matrix1, double[] Matrixele2, int
centerofele)
{
double xCord, yCord;
double MAX=0,distance, tempd1, tempd2;
double[] Distances = new double[Matrix1.length];
int reference, i, j, temp1, temp2, point, length;
reference = i = j = temp1 = temp2 = point =0;
int[] centroids;

class Stack<Item> implements Iterable<Item> {
private final PrintStream StdOut = null;
private int N;              // size of the stack
private Node<Item> first;
private Object StdIn;     // top of stack

// helper linked list class
class Node<Item> {
private Item item;
private Node<Item> next;
```

```java
}

/**
 * Initializes an empty stack.
 */
public Stack() {
first = null;
N = 0;
}

/**
 * Is this stack empty?
 * @return true if this stack is empty; false otherwise
 */
public boolean isEmpty() {
return first == null;
}

/**
 * Returns the number of items in the stack.
 * @return the number of items in the stack
 */
public int size() {
return N;
}

/**
 * Adds the item to this stack.
 * @param item the item to add
 */
public void push(Item item) {
Node<Item> oldfirst = first;
first = new Node<Item>();
first.item = item;
first.next = oldfirst;
N++;
}

/**
 * Removes and returns the item most recently added to this stack.
 * @return the item most recently added
 * @throws java.util.NoSuchElementException if this stack is empty
 */
public Item pop() {
if (isEmpty()) throw new NoSuchElementException("Stack underflow");
Item item = first.item;        // save item to return
```

```java
first = first.next;          // delete first node
N--;
return item;                 // return the saved item
}

/**
 * Returns (but does not remove) the item most recently added to this stack.
 * @return the item most recently added to this stack
 * @throws java.util.NoSuchElementException if this stack is empty
 */
public Item peek() {
if (isEmpty()) throw new NoSuchElementException("Stack underflow");
return first.item;
}

/**
 * Returns a string representation of this stack.
 * @return the sequence of items in the stack in LIFO order, separated by spaces
 */
public String toString() {
StringBuilder s = new StringBuilder();
for (Item item : this)
s.append(item + " ");
return s.toString();
}

public Iterator<Item> iterator() {
return new ListIterator<Item>(first);
}

// an iterator, doesn't implement remove() since it's optional
class ListIterator<Item> implements Iterator<Item> {
private Node<Item> current;

public ListIterator(Node<Item> first) {
current = first;
}
public boolean hasNext()  { return current != null;              }
public void remove()      { throw new UnsupportedOperationException();  }

public Item next() {
if (!hasNext()) throw new NoSuchElementException();
Item item = current.item;
current = current.next;
return item;
```

```java
}
}
public static void main (String args[]) throws IOException
{
Scanner input = new Scanner(System.in);
ClustNumber=input.nextInt();
String [][] numbers = new String [6][2];
double Cordx[] =new double[6];
double Cordy[] =new double[6];
File file = new File("sam.csv");
BufferedReader bufRdr = new BufferedReader(new FileReader(file));
String line = null;
int row = 0;
int col = 0;

//read each line of text file
while((line = bufRdr.readLine()) != null && row< 6 )
{
StringTokenizer st = new StringTokenizer(line,",");
while (st.hasMoreTokens())
{
//get next token and store it in the array
numbers[row][col] = st.nextToken();
col++;
}
col = 0;
row++;
}
for(row=0;row < 6;row++)
{
for(col=0; col<2;col++)
{
System.out.print(" " + numbers[row][col]);
}
System.out.println(" ");}
for(row=0;row<6;row++)
{

Cordx[row]=Double.parseDouble(numbers[row][0]);
Cordy[row]=Double.parseDouble(numbers[row][1]);
//System.out.print(" " + Cordx[row]);
}

for(row=0;row<6;row++)
{
System.out.print(" " + Cordx[row]);
```

```
//System.out.print("\n " + Cordy[row]);
}
System.out.print(" \n");
for(row=0;row<6;row++)
{

System.out.print(" " + Cordy[row]);
}
cent.Grouping(Cordx,Cordy,ClustNumber);

}
}
class StackDemo {
void showpush(Stack st, int a) {
st.push(new Integer(a));
System.out.println("push(" + a + ")");
System.out.println("stack: " + st);
}

/**
 * Unit tests the <tt>Stack</tt> data type.
 */
public void main(String[] args) {
Stack<String> s = new Stack<String>();
while (!((String) StdIn).isEmpty()) {
String item = StdIn.toString();
if (!item.equals("-")) s.push(item);
else if (!s.isEmpty()) StdOut.print(s.pop() + " ");
}
StdOut.println("(" + s.size() + " left on stack)");
}
}
for(j = 1; j < Matrix1.length;j++)
{
Distances[j-1] = Math.sqrt(Math.pow((Matrix1[j]-Matrix1[0]),2) +
Math.pow((Matrixele2[j]-Matrixele2[0]),2));
}

for(i=0;i<Matrix1.length-1;i++)
{
for(j=0;j<Matrix1.length-1-i;j++)
{
if(Distances[j+1] < Distances[j])
{
distance = Distances[j];
```

```
tempd1 = Matrix1[j];
tempd2 = Matrixele2[j];
Distances[j] = Distances[j+1];
Matrix1[j] = Matrix1[j+1];
Matrixele2[j] = Matrixele2[j+1];
Distances[j+1] = distance;
Matrix1[j+1] = tempd1;
Matrixele2[j+1] = tempd2;
}
}
}
//recalculation of centroids
point = Matrix1.length;
do
{
if(Matrix1.length % centerofele != 0)
point--;
}while(point % centerofele != 0);
length = point/centerofele;
for(i=0;i<Matrix1.length;i=length+i)
{
if((i+length-1) > point)
break;
tempd1 = Matrix1[i];
tempd2 = Matrixele2[i];
Matrix1[i] = Matrix1[i+length-1];
Matrixele2[i] = Matrixele2[i+length-1];
Matrix1[i+length-1] = tempd1;
Matrixele2[i+length-1] = tempd2;
}
}
}
```